\documentclass[12pt,preprint]{aastex}

\begin {document}

\title{Evolution of the Inner Circumstellar Envelope of V838 Monocerotis}

\author{J.P. Wisniewski\altaffilmark{1}, K.S. Bjorkman\altaffilmark{1},
A.M. Magalh\~aes\altaffilmark{2}}

\altaffiltext{1}{Ritter Observatory, Department of Physics and Astronomy, University of Toledo, Toledo, OH 43606-3390 USA, jwisnie@physics.utoledo.edu, karen@physics.utoledo.edu}
\altaffiltext{2}{IAG, Universidade de S\~ao Paulo, Caixa Postal 3386, S\~ao Paulo, SP 01060-970, Brazil, mario@astro.iag.usp.br}

\begin{abstract}

We present imaging polarimetry observations of the eruptive variable V838
Monocerotis and its neighboring field obtained in 2002 October.  The
polarization of field stars confirms the previously determined interstellar 
polarization along the line of sight to V838 Mon.  While V838 Mon showed
intrinsic polarization shortly after its second outburst on 2002 February 8,
all subsequent observations only showed a quiescent interstellar polarization
component.  We find V838 Mon
once again showed significant intrinsic polarization in 2002 October,
suggesting the presence of an asymmetrical geometry of
scattering material close to the star.  Furthermore, an
 observed 90$^{\circ}$ position
angle flip
in the intrinsic polarization from 2002 February to 2002 October suggests
that the distribution of nearby circumstellar material has experienced
significant changes.  We discuss the opacity changes in the evolving
circumstellar cloud around V838 Mon that may explain these observations.

\end{abstract}


\section{Introduction}

The eruptive variable V838 Monocerotis, first reported as
 a possible nova
on 2002 January 6.6 \citep{bro02}, experienced three significant photometric
outbursts in early 2002 \citep{mu102,kim02}.  From pre-outburst to its maximum
brightness during the second outburst, V838 Mon brightened by over
nine magnitudes in V, from 15.85 to 6.66 \citep{gor02}.  V838 Mon
also exhibited significant spectroscopic variability in early 2002: neutral
metal and s-process lines were observed following the first outburst 
\citep{zwi02}, ionized metal lines were noted following the second outburst
\citep{iij02,mor02}, and neutral metal and molecular lines were
observed following the third outburst \citep{rau02,ban02}.  V838 Mon exhibited
intrinsic polarization on 2002 February 8 \citep{wis03}.  Polarimetric 
observations after 2002 February 13 only detected the presence of an
interstellar polarization (ISP) component \citep{mu202, wis03}.

V838 Mon developed a light echo by 2002 February 17 \citep{hen02}.   
\citet{bon03} obtained imaging polarimetry of this light echo with HST ACS,
and suggested a distance of 3 to 7 kpc to V838 Mon, although a wide range
 of other distances have been
estimated (e.g. see discussion in Wisniewski et al. 2003 and Tylenda 2003).
  In late September and early October of
2002, V838 Mon's spectral type had evolved to ``later than M10-III'' 
\citep{des02}.  Furthermore, a weak blue continuum was detected by several 
groups, 
suggesting that a binary component of spectral type B3V might be present 
\citep{des02, wag02, mu302}.  Recent MMT spectra confirm this secondary
component (Starrfield, private communication).

Given the magnitude and complexity of the variability exhibited by V838 Mon, it
is not surprising that there is still great uncertainty regarding the 
exact nature of this object.  In this paper, we present imaging polarimetry
observations of V838 Mon and its surrounding field obtained in 2002 October.
These observations allow us to further interpret the evolution of the 
circumstellar matter near V838 Mon since its major outbursts.

\section{Observations}

We obtained polarimetric measurements of V838 Mon and its field 
in 2002 October at the CTIO 1.5 m telescope.  The standard Cassegrain focus
imaging camera with the f/13.5 (0.24 arcsec pixel$^{-1}$) configuration was
 modified by the addition of a
rotatable half-wave plate followed by a fixed analyzer placed before the
second filter wheel.  The analyzer was a double calcite block whose optical
axes had been crossed to minimize astigmatism and color effects, 
which produced two orthogonally polarized images of each object in the field.
One polarization modulation is covered for each 90$^{\circ}$ rotation of the 
waveplate.  For our V, R, and I observations of V838 Mon, we took
images with the waveplate rotated through 8 positions 22.5$^{\circ}$ 
apart.  Further details about this polarimeter can be found in 
\citet{mag96}, \citet{mel01}, \& \citet{per02}.

After basic image processing in IRAF\footnote{IRAF
is distributed by the National Optical Astronomy Observatories, which are
operated by the Association of Universities for Research in Astronomy, Inc.,
under contract with the National Science Foundation.}, we 
performed aperture photometry on the fields.  The reduction
package PCCDPACK \citep{per00}, which calculates linear polarization from a
least squares solution of the photometry in the 8 waveplate positions ($\psi_{i}$), was
then applied.  The residuals at each waveplate position with respect to
the expected $\cos 4\psi_{i}$ curve constitute the uncertainties in our
data; these are consistent with the theoretically expected photon noise
errors \citep{mag84}.  Each object whose polarization is reported has been
carefully checked for contamination effects due to close neighbors; objects
which contained such contamination were excluded from the present study.

Polarized and unpolarized standard stars were monitored nightly throughout
the eleven nights of our observing run, and the stability of the polarimeter 
was reflected in the consistency of these standard star observations.  
Comparing our observations of polarized standard stars to available literature
data, we transformed our polarization measurements to a standard equatorial
system.  This transformation is accurate to a position angle (PA) of better
 than 1$^{\circ}$.  
Instrumental polarization was consistently less than 0.05$\%$.  Note that
the simultaneous observation of the dual orthogonally polarized images of
each target in each waveplate position allows for accurate polarimetry to
be performed even under non-photometric skies, as all sky polarization is
practically canceled \citep{mag96}.  We summarize our observations of
V838 Mon, along with polarimetry of this object from the literature, in 
columns 1-5 of Table 1.

\section{Results}

\subsection{Polarization of Field Stars}

Over small spatial regions, where variations in
the distribution of dust and magnetic fields can be assumed to be
small, field stars located at similar distances exhibit similar levels
of interstellar polarization (ISP).  To study the intrinsic polarization of an
astrophysical object, it is often beneficial to examine the
polarization of spatially nearby field stars, located at similar distances,
 to get a statistical
measure of the ISP along a particular line of sight.

We measured the polarization in V, R, and I of most of the field stars
 surrounding V838
Monocerotis; a sample of these results is given in Table 2 and a complete
listing of these data is available online.  
Figure 1a shows a V band image obtained on 2002 October 16 at the CTIO 0.9m, and
includes numerical references to stars shown in Figure 1b and tabulated in
Table 2.  A plot of the spatial distribution
of the polarization of these field stars in the V band is shown in Figure 1b.
 We note that three stars near V838 Mon show very similar
 polarizations to each other, star \#2, star \#3, and star \#5.  Furthermore,
these stars exhibit very similar polarizations to the ISP previously estimated
for V838 Mon \citep{wis03}.  We plot the wavelength
dependence of the polarization of these three objects in Figure 2, and
overlay the modified Serkowski law \citep{ser75,wil82} description of 
V838 Mon's interstellar polarization with $P_{max} = 2.746 \pm 0.011 \%$,
 $\lambda_{max} = 5790 \pm 37\AA$, $PA = 153.43 \pm 0.12^{\circ}$,
$\delta PA (\lambda) = 0$, and $K = 0.971$ \citep{wis03}.  The good agreement
shown in this figure supports the previously determined ISP, derived entirely
from spectropolarimetry. 
 Furthermore, these results imply that stars \#2, \#3, and \#5
are located at similar distances as V838 Mon.  
Future efforts to determine the distance to V838 Mon should 
ensure consistency with these three neighboring objects.

\subsection{Intrinsic Polarization of V838 Mon}

We subtracted the ISP along the line of sight \citep{wis03} from our
observations of V838 Mon to determine its intrinsic polarization.  The
non-zero results (columns 6-8, Table 1) show that on 2002 Oct 22-24 V838 
Mon once again exhibited a significant intrinsic polarization component.
Interestingly, the magnitude
of intrinsic polarization present on 2002 October 22-24 is similar to that 
observed on 2002 February 8; however, the position angle of this scattered
light differs by roughly 90$^{\circ}$, e.g. a PA ``flip'' has occurred.

The presence of an intrinsic polarization component can be interpreted as
a signature of light scattering off circumstellar material which is
distributed in an asymmetrical geometry.  The intrinsic polarization reported
here and by \citet{wis03} is produced in a region immediately surrounding the
unresolved central star, and as such is different from the reported polarization
of V838 Mon's light echo \citep{bon03}.  In our observations of the unresolved
source, the observed polarization 
is roughly the ratio of the scattered to total (i.e. direct plus scattered) 
light received.   Thus the renewed intrinsic
polarization component suggests that either a) a new source of
asymmetrical circumstellar material, i.e. a new source of scatterers, formed
around V838 Mon in 2002 October; b) the opacity of the circumstellar envelope
surrounding V838 Mon evolved significantly between February 5 to February 13 to
October 22-24, thereby changing the dominant source of scattered light
over time; c) the illuminating source had changed; or d) a combination of
the aforementioned scenarios.

\section{Discussion}

\citet{sch92} interpreted an
observed wavelength-dependent PA flip in polarimetric observations of an
unresolved B[e] star as evidence of a bipolar nebula.  These authors argued
 that at short wavelengths
 an optically thick circumstellar disk blocked out most starlight and thus
little scattered light (with a PA aligned along the polar axis) originated
from the equatorial disk.  Most of the polarized light
at short wavelengths originated from the polar
regions, where the resulting PA would be aligned with the equatorial disk.
At long wavelengths, where the disk was optically thin,
the dominant source of scattered light was the equatorial disk region,
 not the polar regions.  The net effect of such a wavelength dependent
opacity effect was the production of a 90$^{\circ}$ PA flip in the
polarization signal at the wavelength where the equatorial and polar regions
 contributed equal amounts of scattered light.

We suggest that a similar type of scenario might explain the renewed intrinsic
polarization component and PA flip in V838 Mon.  On 2002 February 8, V838
Mon had an intrinsic polarization component, indicating the presence of
an asymmetrical distribution of circumstellar material, initially suggested
to be a flattened circumstellar envelope \citep{wis03}.  We postulate that at
this initial stage, the observed scattered light originated primarily from
specific physical locations, e.g. the polar 
region.  As the opacity of the circumstellar material evolved over time,
specifically as the opacity of the equatorial material decreased, we
suggest that the contribution of scattered light from the polar and equatorial
regions were nearly equivalent.  Thus, viewed as an unresolved object, the
scattered light from the circumstellar envelope would appear to be unpolarized,
consistent with observations in late 2002 February and March 
\citep{wis03,mu202}.  As the envelope continued to
evolve and the equatorial region experienced a further decline in opacity,
one would expect the equatorial region to slowly emerge as the dominant source
of scattered light.  This re-emergence of a dominant (and secondary)
scattering region would produce an intrinsic polarization component 
oriented 90$^{\circ}$ from the original position angle.  The projection of the
intrinsic PA of this envelope onto the sky (2 $\theta \sim 75 ^{\circ}$,
measured N to E) is not inconsistent with the overall morphology seen in the HST
ACS images \citep{bon03}.  The intrinsic PA of the polar region is also not
inconsistent with the direction of the hole in the HST ACS light echo images.

We note that the opposite relative opacity changes would produce a similar 
PA flip.  The scattered light in the circumstellar envelope could have initially
been dominated by an optically thin disk.  As this equatorial material
dispersed, the scattered light from the equatorial and polar regions could
have balanced, producing zero intrinsic polarization.  Over time, 
expansion of the disk could have evacuated the equatorial region, causing
the polar region to emerge as the dominant source of scattered light.

Followup infrared (IR) imaging polarimetry would be valuable in providing
additional constraints on the nature of V838 Mon's circumstellar envelope.
Based upon infrared (IR) light and color curves, \citet{cra03} suggest that
a dust shell had formed around the central star by 2002 April.  If
in early 2002 this dust formed in the equatorial regions as described in the
aforementioned first scenario, i.e. an initial thick equatorial and thin polar
region, then the IR polarization position angle would be the same as that
quoted in this paper.

As described in the introduction of this paper, V838 Mon appears to have
a binary companion.  
 Polarimetric observations of mass transfer binary
systems show some degree of periodicity and scattering at a constant
PA, defined by the orbital plane of the system \citep{hof98}.  The observed
PA flip in V838 Mon's intrinsic polarization, along with the lack of any
apparent periodicity, suggest that binarity is not the dominant source of
scatterers responsible for the observed intrinsic polarization component.

In summary, the polarization of field stars 
confirms the previously suggested interstellar polarization along
the line of sight to V838 Mon.  We find strong evidence of a
 renewed intrinsic polarization component in V838 Mon, 
at a position angle nearly 90$^{\circ}$ from that present on 2002 February 8.
 We suggest that these observations indicate an evolution
of the circumstellar environment of V838 Mon, possibly characterized by:
1) an initially optically thick
equatorial disk which contributed a limited amount of scattered light; 2) a
 subsequent decline in disk opacity resulted in a balance of scattered light
produced by the polar and equatorial regions; and 3) the further decline in the
opacity of the equatorial disk finally allowed the equatorial region to
become the dominant source of scattered light.
  Continued spectropolarimetric monitoring of V838 Mon is strongly
encouraged, as such observations would enable detailed modeling of the
circumstellar environment of this unique object to be performed.

\acknowledgments

We thank the referee, S. Starrfield, for his suggestions to improve this paper.
We also thank Antonio Pereyra and Brian Babler for assisting with the 
data reduction.
This work has been supported in part by NASA LTSA grant NAG5-8054 to the
University of Toledo.  KSB is a Cottrell Scholar of the Research Corporation
and gratefully acknowledges their support.  JPW is supported by a NASA GSRP 
fellowship and thanks NOAO for supporting his travel to CTIO.
AMM acknowledges travel support by FAPESP; he is also partially supported by
 CNPq.  Polarimetry at the University of S\~ao Paulo (USP) is supported by
 FAPESP.  AMM and KSB acknowledge partial travel support by USP.

\clearpage
\newpage

\begin{figure}
\epsscale{0.8}
\plotone{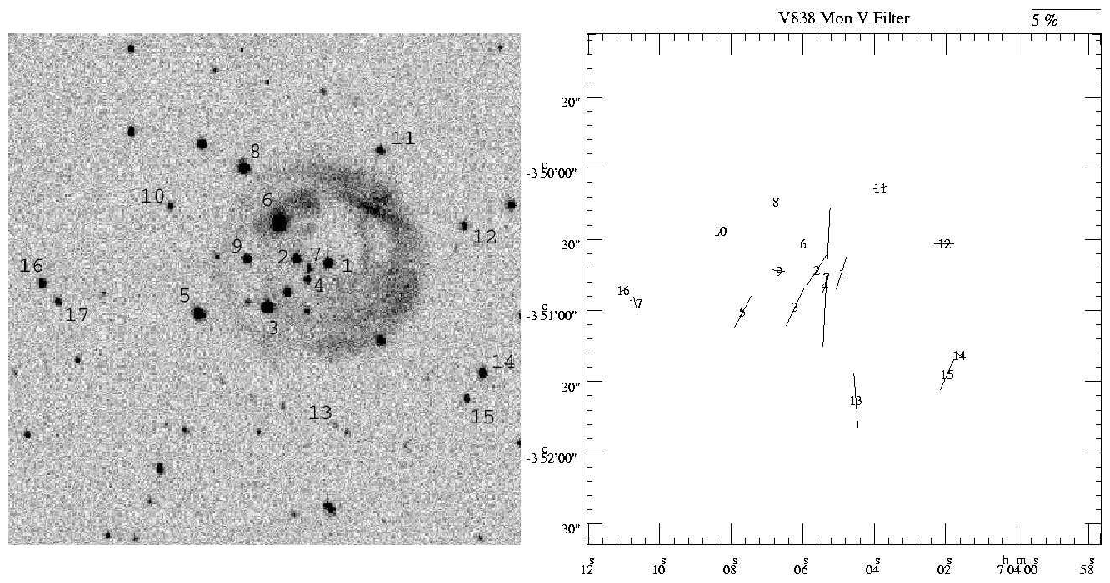}
\figcaption[f1.eps]{Fig 1a: The location of the field stars cited in
 Figure 1b and in Table 2 are overlayed on a V band image of the field
of V838 Mon, taken with the CTIO 0.9m.  V838 Mon is labeled as number 1.
Fig 1b: Polarization of V838 Mon and surrounding field stars.}
\end{figure}

\newpage
\begin{figure}
\epsscale{0.35}
\plotone{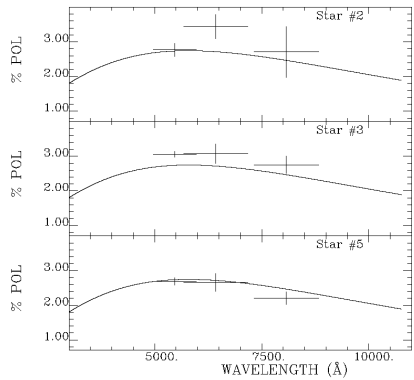}
\figcaption[f2.eps]{The wavelength dependence of three nearby field stars
 of V838 Mon is shown, along with an overlay of the interstellar polarization
 Serkowski fit derived by \citet{wis03} from spectropolarimetry in early 2002.
   These field stars are consistent with
this previously determined ISP.}
\end{figure}

\newpage
\begin{table}
\caption{}
\scriptsize
\begin{tabular}{lcccccccc}

Filter & Date & \%P  & \%Err & PA & Intrinsic \%P & \%Err & Intrinsic PA & Remarks \\

\tableline

B & 2002 Feb 8  & 2.840 & 0.081 & 146.6 & 0.703 & 0.081 & 117.0 & \citet{wis03} \\
V & 2002 Feb 8  & 3.412 & 0.012 & 146.7 & 0.983 & 0.012 & 127.0 & \citet{wis03}\\
V$^{*}$ & 2002 Feb 13 & 2.729 & 0.009 & 153.4 & \nodata & \nodata & \nodata &
\citet{wis03} \\
V & 2002 Oct 24, 08:12UT & 2.49 & 0.12 & 162.3 & 0.90 & 0.12 & 32.0 & this study \\
V & 2002 Nov 12 & 3.05 & 0.35 & 159.7 $\pm$ 3.9 & \nodata & \nodata & \nodata &
\citet{gir02} \\
5500\AA & 2002 Feb-March & 2.6 & \nodata & 150 $\pm$ 2 & \nodata & \nodata & \nodata & \citet{mu202} \\
R & 2002 Feb 8 & 3.226 & 0.004 & 149.0 & 0.714 & 0.004 & 131.2 & \citet{wis03} \\
R & 2002 Feb 13 & 2.667 & 0.004 & 153.4 & \nodata & \nodata & \nodata & \citet{wis03} \\
R & 2002 Oct 22, 08:57UT & 2.24 & 0.21 & 163.2 & 0.93 & 0.21 & 36.7 & this study \\
I & 2002 Feb 8 & 2.910 & 0.003 & 149.5 & 0.578 & 0.003 & 131.9 & \citet{wis03}\\
I & 2002 Feb 13 & 2.458 & 0.003 & 153.5 & \nodata & \nodata & \nodata & \citet{wis03} \\
I & 2002 Oct 24, 08:46UT & 2.23 & 0.04 & 156.3 & 0.32 & 0.04 & 42.1 & this study \\

\tablecomments{Summary of our observations of V838 Monocerotis, as
well as literaure polarimetric observations.  The
$^{*}$ superscript denotes a spectropolarimetric observation which did not
span the entire wavelength range of the Johnson V filter, thus the V band
polarization for this observation is thus only an estimate. 
\citet{wis03} suggested, and the present field star measurements support,
that the polarization of V838 Mon on 2002 February 13 was purely
interstellar in origin.  We determined the intrinsic polarization present on 
2002 October 22-24 by subtracting the observed polarization on 2002 
February 13 from our data.}

\end{tabular}
\end{table}

\clearpage
\newpage
\begin{table}
\caption{}
\scriptsize
\begin{tabular}{lcccccc}

RA(2000) & Dec(2000) & Fig. 1 Ref. & Filter & \%P  & \%Err & PA  \\

\tableline

7:04:05 & -3:50:50  & 1 & V & 2.486 & 0.118 & 162.3 \\
 \nodata & \nodata  & \nodata & R & 2.236 & 0.209 & 163.2 \\
 \nodata & \nodata  & \nodata & I & 2.230 & 0.043 & 156.3 \\

\tablecomments{Filter polarimetry of targets shown in Figure 1a,b. 
The reference numbers in column three refer to the labels in Figure 1a,b.  The
complete version of this table is in the electronic edition of the
Journal.}

\end{tabular}
\end{table}

\end{document}